\documentclass[%
 reprint,
 amsmath,amssymb,
 aps,
]{revtex4-1}

\usepackage{graphicx}% Include figure files
\usepackage{dcolumn}% Align table columns on decimal point
\usepackage{bm}% bold math
\begin{document}

%\preprint{APS/123-QED}

\title{Constraints on primordial black holes from interstellar dust temperature \\
in the Galaxy}% Force line breaks with \\

\author{A.N. Melikhov}
 \email{melikhov94@inbox.ru}
\affiliation{%
P.N. Lebedev Physical Institute, Leninsky prospekt, 53, Moscow 119991, Russia}%

\author{E.V. Mikheeva}
 \email{helen@asc.rssi.ru}
\affiliation{
Astro Space Center of the P.N. Lebedev Physical Institute, Profsoyusnaya 84/32, Moscow, Russia 117997
}%

\date{\today}% It is always \today, today,

\begin{abstract}
The photons emitted by primordial black holes (PBHs) can be a heating factor for interstellar dust. Assuming that dust in the Galaxy has a homogeneous distribution and that PBHs have the same distribution as dark matter (DM), we constrain the fraction of PBHs in DM from the observational dust temperature in the Galaxy.

\end{abstract}

\maketitle

%\tableofcontents
\noindent {\small Physical Review D \textbf{107}, 063535 (2023)}

\noindent {\small DOI: 10.1103/PhysRevD.107.063535}

%%%%%%%%%%%%%%%%%%%%%%
\section{Introduction}
%%%%%%%%%%%%%%%%%%%%%%
Despite several decades of searches, the dark matter (DM) particle still remains elusive \cite{alex2021}. This fact and the observational discovery of gravity waves emitted by merging black holes \cite{abbott16} has increased interest in primordial black holes (PBHs) as a DM candidate.

Unlike astrophysical BHs, PBHs can have almost arbitrary masses. This means that a phenomenological approach, involving consideration of different mass functions, looks preferable to direct testing of physical models in which PBHs can be created.

Depending on the mass, PBH can manifest itself in different ways. PBHs with mass $M < 10^{15}$ g evaporated to the present could have influenced the processes in the early Universe, such as primordial nucleosintesis \cite{carr10}, and bariogenesis \cite{turner79, barrow91, upadhyay99, bugaev03}. They could also provide additional source of neutrino \cite{bugaev03, bugaev02}, gravitinos \cite{khlopov06}, or other particles \cite{green99,lemoine00}, could swallow monopoles \cite{izawa84, stojkovic05}, remove domain walls \cite{stojkovic205}, and reionise the Universe \cite{he02, mack08}.

If PBH with masses $10^{13}-10^{18}$ g could a significant contribution to DM, then their time-integrated Hawking radiation should make a significant contribution to the observed gamma-rays and X-ray background \cite{carr20, carr10, arbey20, ballesteros20, iguaz21}.

PBHs with $M>10^{15}$~g might influence the development of large-scale structure \cite{meszaros275, carr77, freese83, carr83, afshordi03}, seed the supermassive black holes \cite{carr84, duchting04, khlopov05, bean02}, generate background gravitational waves \cite{carr80, nakanura97, ioka99, inoue03, hayasaki09}, or produce X-rays through accretion and thereby affect the thermal history of the Universe \cite{ricotti08}.

Various physical effects can be used to constrain PBHs, such as direct measurement of $\gamma$-ray photons emitted by PBHs, gravitational lensing, or recently proposed measuring temperature of interstellar gas heated by PBH emitted electrons \cite{kim2021}.  In this paper we adopted a novel method to constrain a number of PBHs in the Universe. First presented in \cite{melikhov} the method takes into account the infrared radiation emitted by dust grains irradiated by photons of Hawking radiation.

%%%%%%%%%%%%%%%%%%%%%%%%%%%%%
\section{Heating and cooling}
%%%%%%%%%%%%%%%%%%%%%%%%%%%%%
Interstellar dust is a component of the interstellar me\-dium. It has been forming in the atmospheres of stars, interstellar clouds, and planetary and protostellar nebulae. Dust grains are transported into the interstellar medium under the action of gas flows and radiation pressure \cite{bochkarev}.

An important property of interstellar dust is that it absorbs and scatters radiation, which leads to reddening and dimming of starlight \cite{tielens}. After absorbing a photon, the dust grain emits infrared radiation in a continuous spectrum, which can be approximated by the Planck spectrum. 

In the paper we accept the widely used Mathis-Rumpl-Nordsieck (MRN) model of interstellar dust \cite{mathis77}. According to this model, interstellar dust grains are spherical in shape and consist of a mixture of graphite and silicate particles in approximately equal mass proportions. Grain sizes lie in the interval $0.005<a<0.25$ $\mu$m with power-law distribution $n(a)\sim a^{-3.5}$. The MRN model explains the interstellar extinction curve in the wavelength range $1100\!-\!10000\, {\buildrel _{\circ} \over {\mathrm{A}}}$ (see Fig.~1 in \cite{mathis77}).

We assume that photons propagate freely, and therefore the interaction with matter can be neglected. Radiation from other sources is not taken into account when calculating the dust heating rate. The equilibrium dust temperature $T_d$ is determined from the heat balance condition. We calculate dust heating grains by PBH photons and dust cooling grains separately for the graphite and silicate components of the dust. Comparing the heating rate and the cooling rate of dust under the assumption that the heating rate has to be less than the cooling rate, we place an upper bound on the PBH dark matter fraction.

%%%%%%%%%%%%%%%%%%%%
\subsection{Heating}
%%%%%%%%%%%%%%%%%%%%
According to \cite{hawking74, hawking75}, the PBH radiation temperature, $T$, is given by
%(1)
\begin{equation}
    k_BT = \frac{\hbar c^3}{8\pi GM}, 
\end{equation}
where $k_B$ the Boltzmann constant, and $M$ is the BH mass.

The photon production rate per intrinsic degree of freedom per unit volume is equal to
%(2)
\begin{equation}
    \frac{dN_{\gamma}}{dt\,dE} = \frac{\Gamma}{2 \pi\hbar}\Big[\exp\Big(E/k_BT\Big)-1\Big]^{-1},
\end{equation}
where $\Gamma$ is the absorption probability of emitted particles by a BH. It denotes the greybody factor which describes the deviation of BH spectrum from the blackbody one. For high energies $\Gamma=\dfrac{27G^2M^2E^2}{\hbar^2c^6}$; see \cite{macgibbon90}.

The heating rate of dust grain is determinated by 
%(3)
\begin{equation}
     \dfrac{dE^{abs}}{dt}=4\pi{\sigma}_d\int_0^{\infty}Q(\lambda)J(\lambda)d\lambda,
\end{equation}
where $J(\lambda)$ is the intensity of radiation from all PBHs, $\sigma_d$ is the geometric cross section of dust grain, $Q$ is the absorption efficiency defined as following (see \cite{tielens})
%(4)
\begin{equation}
    Q(\lambda) =
    \begin{cases}
    1, &\text{$\lambda \leq 2\pi a$}\\
    \frac{2\pi a}{\lambda}, &\text{$\lambda > 2\pi a$},
    \end{cases}
\end{equation}
where $a$ is the radius of the dust grain.

To derive the constrains on PBH fraction we considered different mass functions. The first one is monochromatic mass function ($\delta$ function) and the second one is a log-normal distribution. First proposed in \cite{dolgov93}, the latter has the probability function as follows \cite{krishnamoorthy}: 
%(5)
\begin{equation}
g(M) = \frac{1}{\sqrt{2\pi} \sigma M} \exp\Big(\frac{-\log^2(M/\mu)}{2\sigma^2}\Big),
\end{equation}
where $\mu$ and $\sigma$ are constants.

We assume that PBHs is distributed in Galaxy in the same way as DM and use the spherically symmetric Navarro-Frenk-White (NFW) profile \cite{navarro97}
%(6)
\begin{equation}
    \rho(r)=\dfrac{\rho_{0}}{\frac{r}{r_h}\left(1+\frac{r}{r_h}\right)^2},
\end{equation}
where $\rho_0$ and $r_h$ are constants, which are equal to $8.54\times 10^{-3} M_\odot pc^{-3}$ and 19.6 kpc, respectively (see \cite{mcmillan17}). For this density profile the mass of a DM halo inside a sphere with the radius 250 kpc is equal to $1.37\times 10^{12} M_\odot$. It is rather close to the observational value of the Galaxy mass. Further, we denote this radius value as $R_G$.  

As we assume that dust grains are homogeneously distributed in the DM halo, we need a related coordinate system for grains and PBHs. Let $x$, $y$, $z$ be the Cartesian coordinates of PBHs in coordinate system of the Galaxy, $x_d$, $y_d$, $z_d$ be coordinates of a dust grain in the coordinate system of the Galaxy, and $x_p$, $y_p$, $z_p$ be the coordinates of a PBH in coordinate system of the dust grain.

In this case the relationship between coordinates is defined by the evident expression
%(7)
\begin{equation}
    \begin{cases}
    x = x_d + x_p\\
    y = y_d + y_p\\
    z = z_d + z_p.
    \end{cases}
\end{equation}

The distance between a dust grain and PBH in spherical coordinate system of the dust grain is equal to
%(8)
\begin{equation}
\begin{split}
     R^2 =& (r\sin\theta\cos\phi - r_d\sin\theta_d\cos\phi_d)^2 \\
     &+ (r\sin\theta\sin\phi - r_d\sin\theta_d\sin\phi_d)^2 \\
     &+(r\cos\theta - r_d\cos\theta_d)^2,
     \end{split}
\end{equation}
where $r$, $\theta$, $\phi$ are the spherical coordinates of a PBH in coordinate system related with the Galaxy, and $r_d$, $\theta_d$, $\phi_d$ are spherical coordinates of the dust grain in the same coordinate system. 
  
Energy flux from PBHs is equal,
%(9)
\begin{equation}
\begin{split}
    F_{PBH} &= f\frac{n_d}{N_d}\int_{0}^{R_G}\rho(r)r^2\,dr\int_{0}^{\pi}\sin\theta \,d\theta\int_{0}^{2\pi}d\phi  \\  
    &\times\int_{0}^{R_G}{r_d}^2\,dr_d\int_{0}^{\pi}\sin\theta_d\,d\theta_d\int_{0}^{2\pi}\frac{1}{R^2}\,d\phi_d  \\ 
    &\times \int_{M_{min}}^{\infty}\frac{g(M)}{M}\,dM 
    \int_{0}^{\infty}E\frac{dN_{\gamma}}{dt\,dE}\,dE,
    \end{split}
\end{equation}
where $f$ is the dark matter fraction of PBHs, $n_d$ is the number density of interstellar dust, $N_d$ is the total number of dust grains, $R_G$ is the dark matter halo radius, as discribed in notes to eq.~(6), $M_{min}$ is set at $10^{15}$ g since PBHs with lower masses have evaporated by the present, and $E$ is the photon energy emitted by PBH.  We integrate over the positions of the dust grains due to the lack of thermodynamic equilibrium and in order to take into account differences in the distributions of PBHs and dust grains.

When evaporating PBHs with masses $10^{15}-10^{18}$ g, an energy in the range $0.1-100$ MeV is released. For this energy range the wavelengths of PBH radiation is $\lambda_{PBH}\ll2\pi a$, and therefore we can put $Q(\lambda)=1$.

Substituting (4) and (9) in (3) we derive
%(10)
\begin{equation}
     \dfrac{dE^{abs}}{dt}=4\pi{\sigma}_dF_{PBH}.
\end{equation}

%%%%%%%%%%%%%%%%%%%%
\subsection{Cooling}
%%%%%%%%%%%%%%%%%%%%
Dust grains radiate as modified blackbody. The cooling rate of a dust grain is determinated by
%(11)
\begin{equation}
    \dfrac{dE^{rad}}{dt}=4\pi{\sigma}_d\int_0^{\infty}Q(\lambda)B(T_d,\lambda)d\lambda,
\end{equation}
where $T_d$ is the equilibrium temperature of the dust grain, $B(T_d, \lambda)$ is the Planck function \cite{tielens}.

Eqs.~(3) and (11) are valid only for large size particles ($a\geq0.01$ $\mu$m) \cite{bochkarev}. Small grains have a low heat capacity, which leads to a sharp increase in temperature of small dust grains even when small portion of energy is absorbed, and therefore the dust temperature changes abruptly. Between temperature jumps, the majority of small particles of dust cools down to the temperature of the cosmic microwave background radiation. The emission of dust grains occurs mainly when $T_d$ is higher than the equilibrium value \cite{bochkarev}. Therefore, in our model, we will consider only large dust grains with sizes between 0.01 and 0.25 $\mu$m.

The temperature of interstellar dust varies in different parts of the Galaxy. If interstellar dust is located far away from circumstellar shells, in clouds of atomic and molecular hydrogen, the temperature of the dust grains can decrease to $10-20$~K. In HII zones the temperature of the dust grain can be $30-200$~K. In circumstellar envelopes the temperature is at its maximum and can reach $1000-1500$~K \cite{bochkarev}. In dense clouds, the temperature of interstellar dust can be $6$~K \cite{tielens}.

To constrain the fraction of PBHs in DM we consider the following values of dust temperature for silicate 
%(12)
\begin{equation}
    T_{sil}=13.6\Big(\frac{1 \: \mu m}{a}\Big)^{0.06}\,K,
\end{equation}
%(13)
and graphite components
\begin{equation}
    T_{gra}=15.8\Big(\frac{1 \: \mu m}{a}\Big)^{0.06}\,K.
\end{equation}
Eq.~(12) and (13) are obtained considering that the dust is heated by diffuse Galactic interstellar radiation field and CMB; see \cite{tielens}.

When cooling, the dust grain radiates in the infrared range, so for it $\lambda_d > 2\pi a$, and to calculate the cooling rate we take $Q(\lambda) = 2\pi/\lambda$. Substituting $Q(\lambda)$ into (14), we find the dust grain cooling rate,
%(14)
\begin{equation}
\frac{dE^{rad}}{dt} = 4\pi \sigma_d F_d,
\end{equation}
where $F_d$ is flux emitted by a dust grain. 

%%%%%%%%%%%%%%%%%
\section{Results}
%%%%%%%%%%%%%%%%%
The strongest constraints can be derived for the mininal size of dust grain $a=0.01\,\mu m$. For this particle size the temperatures of graphite and silicate dust components are $T_{gra} = 17.93$~K and $T_{sil}=20.83$~K, respectively. In result, we have got the following heating rate:
%(15)
\begin{equation}
\frac{dE^{abs}}{dt}=1.14\times10^{-10}\times f\times\Big(\frac{10^{15}\,g}{M}\Big)^3 \, erg/s.
\end{equation}

In the calculations $a$ enters the expression for $f$ as $\sim a^{0.7}$. Therefore, if we take the maximum value for dust in MRN model, $a=0.25\,\mu$m, then $f$ will increase by the value $\sim10$.

The cooling rate of the silicate dust grain is equal to
%(16)
\begin{equation}
\Big(\frac{dE^{rad}}{dt}\Big)_{sil}=2.76\times10^{-14} \: erg/s,
\end{equation}
and for graphite component is equal to
%(17)
\begin{equation}
\Big(\frac{dE^{rad}}{dt}\Big)_{gra}=5.63\times10^{-14} \: erg/s.
\end{equation}

To constrain the fraction of PBHs we compared the heating rate and the cooling rate of dust, assuming that the former should be less than the latter. In practice, they are equal. But we do not take into account heating from other sources, such as radiation of stars. This is done in order to put an upper limit on the number of PBHs.

Fig. 1 shows the result for the monochromatic PBH mass function. Solid and dotted lines correspond to the upper limit on the PBH fraction, for silicate and graphite grains, respectinely. We compare the constraints with those derived from observational data on Galactic and extragalactic $\gamma$-ray fluxes mentioned in Ref. \cite{carr10, boudaud19, laha20, laha19}. Our constraints are weaker than those derived in these papers.

%============================================fig1
\begin{figure*}
	\includegraphics[scale=1.0]{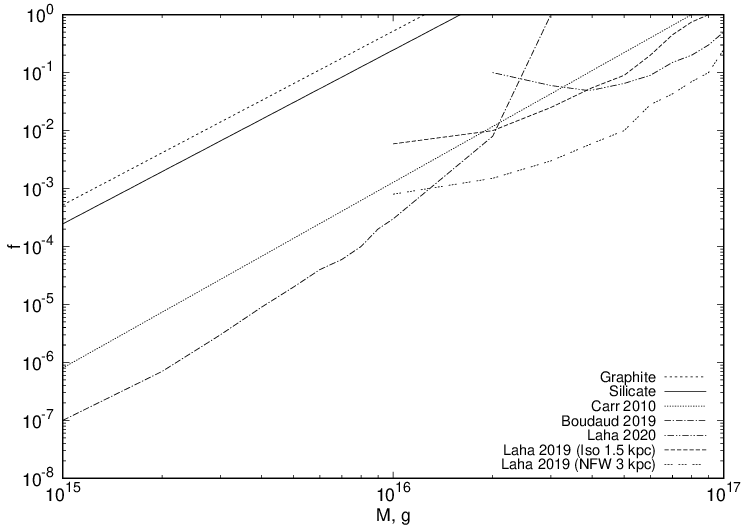}
	\caption{The constraints on a PBH fraction for a monochromatic mass function. The solid (dotted) line shows the constrants if the radiation from the PBHs is absorbed by the silicate (graphite) component of the dust. Other lines correspond to constraints obtained in \cite{carr10, boudaud19, laha20, laha19}; see legend.}
\end{figure*}
%=============================================

\begin{figure*}
	\includegraphics[scale=1.0]{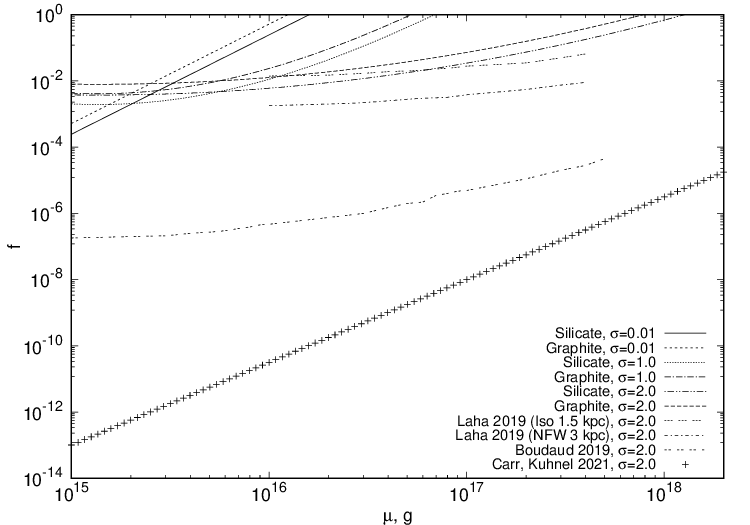}
	\caption{The constraints on a PBH fraction for the case of log-normal mass function obtained for graphite and silicate component of the dust for different values of $\sigma$. Other lines correspond to constraints obtained in \cite{laha19, boudaud19, carr21} for the log-normal distribution with $\sigma=2$; see legend.}
\label{f2}\end{figure*}

Fig.~\ref{f2} demonstrates the results for the log-normal mass distribution. The horizontal axis shows the values of the $\mu$ parameter. Other constraints derived from observational data on Galactic and extragalactic $\gamma$-rays fluxes are presented  here; see \cite{laha19, boudaud19, carr21}. If $\sigma$=0.01 and $\sigma$=1 our constraints are weaker than those derived in these papers. For the case $\sigma$=2 our constraints are roughly similar on those derived in \cite{laha19}, but weaker than those derived in \cite{boudaud19, carr21}.

%=========================================fig2
%===========================================

%%%%%%%%%%%%%%%%%%%%
\section{Conclusion}
%%%%%%%%%%%%%%%%%%%%
In this paper we considered dust heating by PBHs, assuming that dust grains are homogeneously distributed in the Galaxy and PBHs have the same distribution as dark matter. We use NFW profile for the Galactic DM density. As mass functions we regarded the monochromatic mass function and a log-normal distribution. Recently, in \cite{carr17}, a new approach to analyze costraints on PBH fraction in CDM was proposed. However, in the article, we present the constraints for the monochromatic and log-normal mass function separately in traditional way, which looks more explicit.

We calculated heating rate and cooling rate of the dust and derived a constraint on PBH fraction $f$, assuming that heating rate is less than cooling one. In case of a monochromatic mass function the constraints are weaker than those derived in previous papers. For log-normal distribution, when $\sigma$=2, the constraints are roughly similar on those derived in \cite{laha19} and weaker than those derived in \cite{boudaud19, carr21}.

%%%%%%%%%%%%%%%%%%%%%%%%%%
\section*{Acknowledgments}
%%%%%%%%%%%%%%%%%%%%%%%%%%
The authors express their gratitude to the referees for their remarks and comments, which clarified the text of the article.

%%%%%%%%%%%%%%%%%%%%%%%%%%%
\bibliography{refs}% Produces the bibliography via BibTeX.

\end{document}